\documentclass[twocolumn,prl]{revtex4}
\usepackage{graphics,graphicx}
\begin{document}
\title{Nonconventional magnetic phenomena in neodymium thin film}
\author{G. Yumnam$^{1}$}
\author{J. Guo$^{1}$}
\author{Y. Chen$^{1}$}
\author{V. Lauter$^{2}$}
\author{D. K.~Singh$^{1}$}
\affiliation{$^{1}$Department of Physics and Astronomy, University of Missouri, Columbia, MO 65211}
\affiliation{$^{6}$Neutron Scattering Division, Oak Ridge National Laboratory, Oak Ridge, TN 37830}
\affiliation{$^{*}$email: singhdk@missouri.edu}

\begin{abstract}
Neodymium is a remarkable active component in numerous magnetic alloys that are used in various applications. However, the application of bare neodymium thin film is limited due to the lack of information about its electrical and magnetic properties. We report synergistic study of Nd thin film using experimental and theoretical techniques of polarized neutron reflectometry, magnetoresistance measurement and density functional theory. Unlike bulk Nd, thin film specimen is a very poor electrical conductor. Also, as grown thin film on silicon substrate does not exhibit  any magnetism in zero field. However, moderate inplane field application of $H$ = 1.2 T tends to induce weak magnetism in the system at low temperature of $T$ $<$ 18 K, which coincides with an unusual cross-over behavior in magnetoresistance. The study provides important insight in the physical characteristics of Nd thin film that are atypical for a magnetic system. \end{abstract}

\maketitle

Neodymium magnet's prosaic use includes a long list of applications.\cite{Zepf} Some of the examples include uses in the alloys to make magnet for the household purposes, hard disk drive for data storage, magnetic resonance imaging machine and colored glasses and light bulbs.\cite{Herbst,Kenya,Kawamura,Chinn,Sprecher,Lin} Majority of the applications hinge on understanding the physical and magnetic characteristics of neodymium (Nd) in bulk and thin film specimens.\cite{Kemp,Taylor} Besides the practical applications, Nd element is an integral component in the numerous scientific investigations of fundamental importance in solid state and materials physics. For instance, Nd-based compounds are at the forefront of the study of quantum magnetism or the exploration of plasmonic properties.\cite{Sala,Xu,Zhang,Wang,Hatnean,Benton} More recently, an emergent nanomaterial with honeycomb motif, created using Nd element, was used to demonstrate a highly novel state of the Wigner crystal of magnetic charges.\cite{Chen} In the bulk form, neodymium is argued to manifest antiferromagnetic spin density wave type order below $T_N$$\simeq$ 19 K.\cite{Bak,Moon} However, little or no information is available about neodymium thin film; especially electrical and magnetic properties that form the core of physical understanding, necessary for the practical applications. Here, we report on a comprehensive study of neodymium thin film using the synergistic experimental and theoretical investigations. Experimental researches, performed using electrical, magnetoresistance and polarized neutron reflectometry measurements, are complemented by the first principle theoretical calculations under the ambit of the density functional theory formalism. Unlike bulk Nd, thin film of Nd is found to exhibit very poor electrical conduction property. Interestingly, a cross-over behavior between negative and positive magnetoresistance is found to occur at $T \simeq$ 18 K. Inplane magnetic field application tends to induce a weak magnetic order at low temperature, which is otherwise absent in the thin film sample. 

Thin film of neodymium, thickness $\simeq$ 20 nm, is created using electron-beam deposition of neodymium vapor on silicon substrate in an ultra-high vacuum chamber. For e-beam deposition, the high purity (99.9\% or higher) Nd pallets were commercially obtained from Kurt J. Lesker. Four probe technique was employed to measure the electrical resistance and magnetoresistance of Nd thin film. Electrical measurements were performed in a closed cycle refrigerator cooled 9T magnet with the measurement temperature range of $T$  = 5 K - 300 K. Due to the high electrical resistance, electrical measurements were carried out using the synchronized combination of Keithley 6221 current source meter and 2182 nanovoltmeter in the delta mode configuration. Polarized neutron reflectometry (PNR) measurements were performed on a 20$\times$20 mm$^{2}$ surface area sample at the Magnetism Reflectometer at beam line BL-4A of the Spallation Neutron Source (SNS) at the Oak Ridge National Laboratory.\cite{VL} The instrument operates using the time of flight technique in a horizontal scattering geometry with a bandwidth of 5.6 $\AA$ (wavelength varying between 2.6 - 8.2 $\AA$). The beam was collimated using a set of slits before the sample and measured with a 2D position sensitive $^{3}$He detector with 1.5 mm resolution at 2.5 m from the sample. The sample was mounted on the copper cold finger of a close cycle refrigerator with a base temperature of $T$ = 5 K. Beam polarization was performed using the reflective super-mirror devices, achieving better than 98 \% polarization efficiency over the full wavelength band. Also, the full vertical divergence was used for maximum intensity and a 5\% angular relative resolution $\Delta$$\theta$/$\theta$ $\simeq$ $\Delta$q$_{z}$/ q$_{z}$ in the horizontal direction. Measurement in magnetic field utilized an electromagnet with a maximum field of $H$ = 1.2 T, applied inplane to the sample.

\begin{figure}
\centering
\includegraphics[width=8.9 cm]{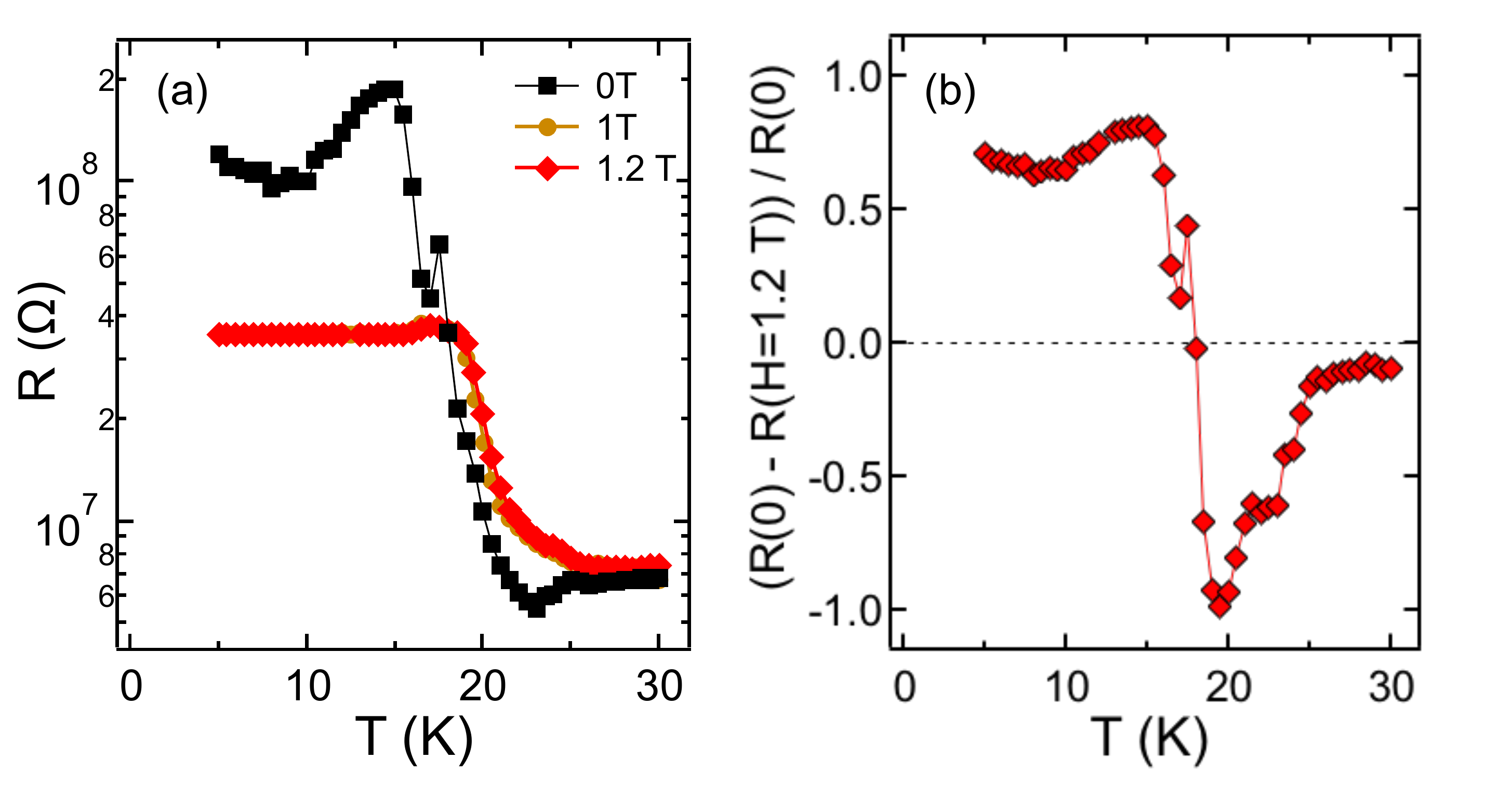} \vspace{-6mm}
\caption{(color online) Electrical characterization of Nd thin film. (a) Electrical resistance as a function of temperature at $H$ = 0 T and $H$ = 1.2 T. A clear separation between the two resistance curves occur at $T$ $\simeq$ 25 K. Below $T$ $\simeq$ 18 K, zero field resistance exceeds the field data. (b) MR vs. temperature. Remarkable crossover from negative to positive MR occurs around $T$ $\simeq$ 18 K.
} \vspace{-6mm}
\end{figure}

We plot the electrical resistance of Nd film as a function of temperature at $H$ = 0 T and $H$ = 1.2 T field in Fig. 1a. The highly resistive nature of the film becomes immediately noticeable. As temperature reduces, a steep rise in resistance is detected below $T \simeq$ 25 K in $H$ = 0 T field data. However, the resistivity tends to flatten for further decrease in temperature below $T \simeq$ 15 K (albeit, preceded by a spike). Although, the sharp upward trend in electrical resistance is usually ascribed to the thermally activated behavior, the flattening characteristic at further lower temperature suggests a competing intrinsic mechanism at play. This argument is further substantiated by comparing the $H$ = 0 T measurement with $H$ = 1.2 T data in Fig. 1a. Three important observations can be inferred here: (a) the resistivity curve at $H$ = 1.2 T separates out from the $H$ = 0 curve below T = 25 K, suggesting the onset of magnetism below $T$ $\simeq$ 25 K in Nd thin film, (b) resistivity curves tend to flatten below $T$ $\simeq$ 20 K ($H$ = 1.2 T) or $T$ $\simeq$ 15 K ($H$ = 0 T), and (c) R($H$) is significantly smaller than R(0) below $T$ $\simeq$ 18 K. The reduction in electrical resistance below $T$ $\simeq$ 18 K in applied field is most likely arising due to the field-induced alignment of magnetic moment along the field direction; thus, causing a ferromagnetic-type moment arrangement in the system. Ferromagnetic configuration is known to give rise to lower resistance in a magnetic materials.\cite{Zeng,Berger} We further understand it by plotting magnetoresistance (MR), given by (R(0) - R($H$=1.2 T))/R(0), as a function of temperature in Fig. 1b. The plot of MR vs. temperature exhibits a very interesting cross-over behavior from -ve values, at 18 K $< T < $25 K, to +ve values at $T$ $<$ 18 K. Negative magnetoresistance is often associated to the variable range hopping mechanism of charge carriers between the exchange coupled magnetic clusters in a magnetic material.\cite{Parker,Gunasekera} Therefore, electrical conduction, in the temperature range of 18 K $< T <$ 25 K, may be occurring due to the hopping of charge carriers. On the other hand, the positive value of MR is argued to be arising due to the thermally activated mechanism in magnetic systems.\cite{Gunasekera2} However, this argument is not applicable to electrical resistance temperature dependence below $T$ $\simeq$ 18 K in Nd thin film, as it does not manifest exponential rise as a function of (reducing) temperature. Rather, resistance flattens at lower temperature. At $T$ $<$ 18 K, the system tends to develops weak field-induced magnetism. 

\begin{figure}
\centering
\includegraphics[width=9.0 cm]{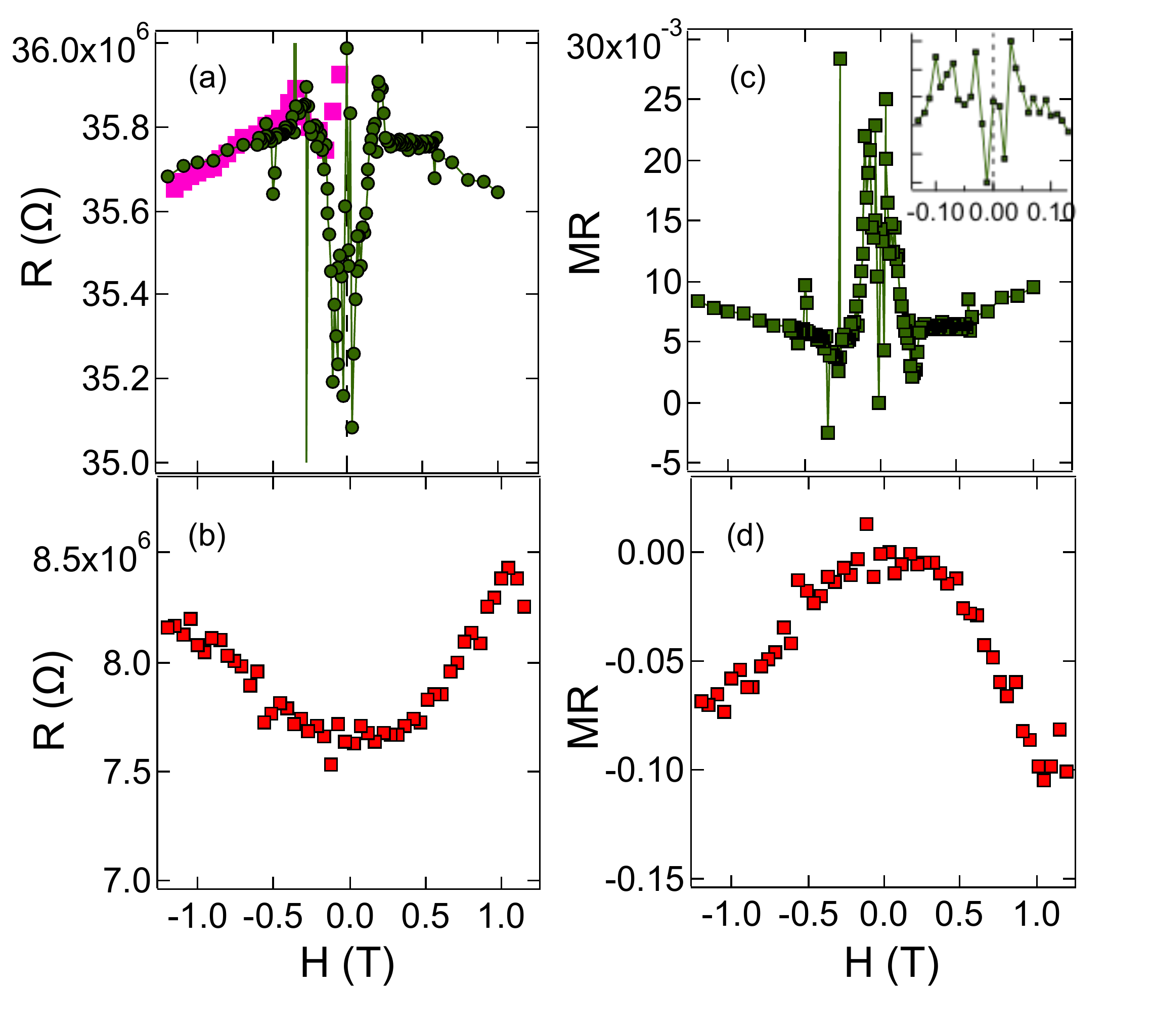} \vspace{-8mm}
\caption{(color online) Magnetoresistance measurement of Nd film. (a-b) $R$ vs. $H$ at $T$ = 5 K and $T$ = 30 K. Unlike the resistance vs. field at $T$ = 30 K, complex behavior is observed near zero field in $T$ = 5 K data. Pink color marker is from the zero field cooled measurement at $T$ = 5 K, immediately after cooling the sample from $T$ = 300 K to $T$ = 5 K in zero magnetic field. (c-d) MR, given by (R(0) - R($H$))/R(0), vs. $H$ at $T$ = 5 K and $T$ = 30 K. Resistance in zero field cool data at $H$ = 0 T is used as $R$(0), which is higher than the value in field sweep mode. MR data at different temperatures exhibit opposite trends in field e.g. MR at $T$ = 30 K exhibits linear decrement, typically observed in a paramagnetic system, as magnetic field increases, while the MR at $T$ = 5 K increases as a function of field at $H >$ 400 Oe. At low field, highly complex trends in MR, indicating the existence of quasi-stable states at $T$ = 5 K, are detected (see inset in Fig.c).} \vspace{-6mm}
\end{figure}

\begin{figure*}
\centering
\includegraphics[width=16 cm]{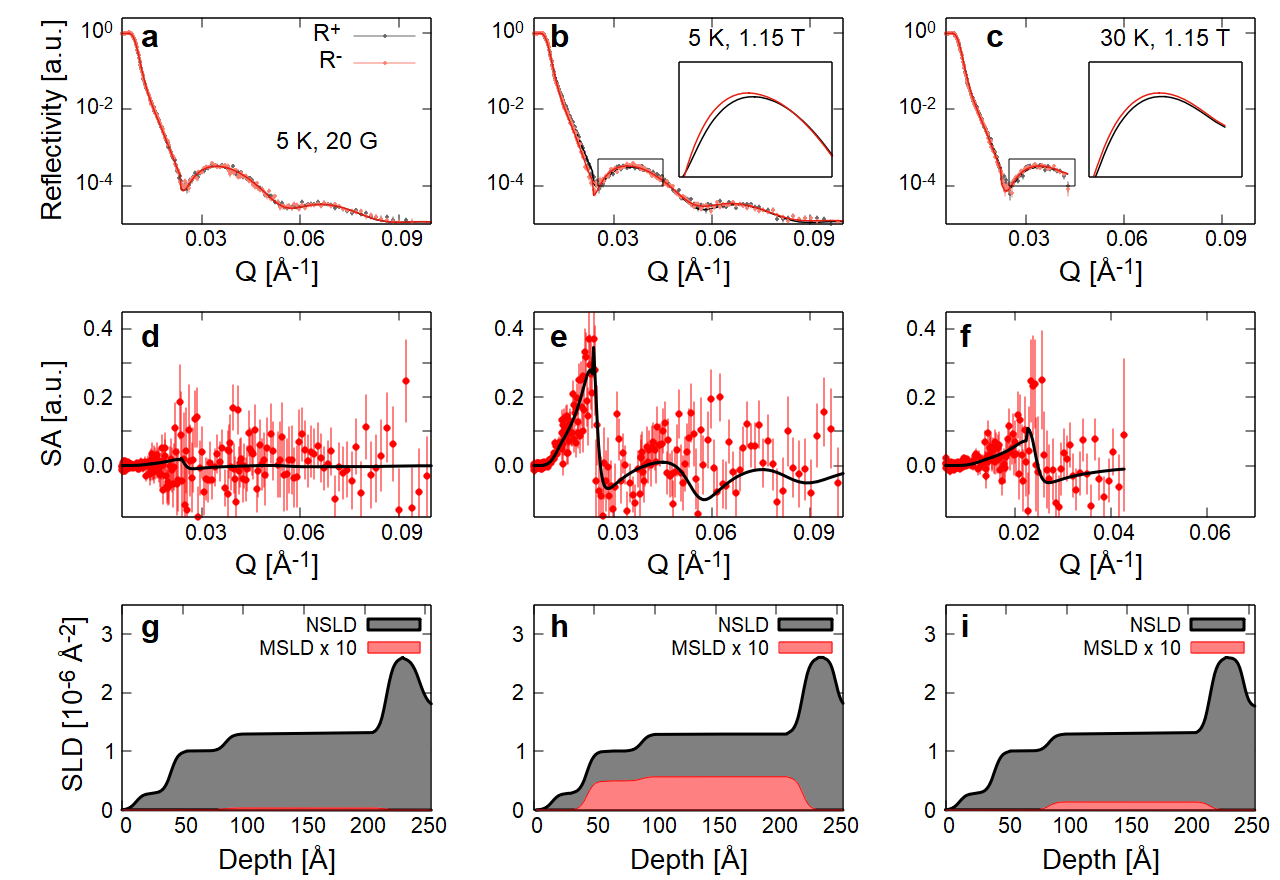} \vspace{-2mm}
\caption{(color online) Polarized neutron reflectometry measurements on Nd thin film. (a-c) A small guide field of $H$ = 20 Oe is applied to keep neutron polarized. Measured and fitted (solid curves) reflectivity curves for neutron with spin-up (R$^{+}$) and spin-down (R$^{-}$) polarizations as a function of wave vector transfer at different temperatures and magnetic fields. (d-f) Nuclear and magnetic scattering length density profiles, obtained from the fit to experimental data, as a function of depth of Nd thin film. (g-i) Plot of spin asymmetry SA at different temperatures and fields, obtained from the experimental and fitted reflectivities in (a-c). A very small MSLD is extracted in the data at $T$ = 5 K at $H$ = 1.2 T. No magnetism is inferred to arise in zero magnetic field.
} \vspace{-4mm}
\end{figure*}

The quantitative information about field-induced magnetism in Nd thin film can be obtained from analyzing the magnetoresistance data as a function of applied magnetic field. MR data as a function of magnetic field were acquired at two temperatures of $T$ = 5 K (in the field-induced magnetism regime) and $T$ = 30 K (above the magnetism onset temperature). For MR measurement purposes, the sample was cooled from $T$ = 300 K to $T$ = 5 K in zero field. Plots of magnetoresistance as a function of magnetic field in Nd thin film are shown in Fig. 2. Distinct electrical responses to applied field are observed in $T$ =5 K and $T$ = 30 K data. While a uniform linear increment is observed in negative MR as a function of magnetic field at $T$ = 30 K, measurements at $T$ = 5K reveal complex behavior. The linear variation in negative MR at $T$ = 30 K is a typical characteristic of paramagnetic system.\cite{Negisi} The thin film behaves as a paramagnet at higher temperature of $T >$ 25 K. An entirely different picture emerges in the MR measurement at low temperature. In an unusual observation, MR first decreases to a near zero value, $\simeq$ 0.002, at low field of $H$ = 150 Oe, before quickly bouncing back by an order of magnitude to the highest MR of 0.023 at $H$ = 300 Oe (see inset in Fig. 2c). It seems that the system undergoes through quasi-stable states near zero field, because the zero field cool magnetoresistance measurement, performed immediately after the sample was cooled from $T$ = 300 K to 5 K, also exhibits a small dip in resistivity near the zero field. When magnetic field is swept from the negative value to the positive value (and vice-versa), a large changes in MR is observed near the zero field. It could be arising due to a competing mechanism where the system tries to achieve the zero-field state and thus minimize the energy, but the remnant magnetization in the sample due to magnetic field application inhibits that tendency.\cite{Berger,Levy} As a result, the Nd film develops quasi-stable configuration near zero field. Following the unusual back and forth bounces between the high and low MR near zero field, sanity returns as field increases further. The MR exhibits a gradual increment at higher field. However, the overall magnetoresistance is very small, 0.01 at $H$ = 1.2 T, which suggests the existence of weak field-induced magnetism in the system.

A more direct evidence to the field-induced magnetism in Nd thin film is obtained from the polarized neutron reflectometry (PNR) measurements.\cite{Valeria} PNR measurements were performed in a small guide field of $H$ = 20 Oe to maintain the polarization of incident and scattered neutron. Experimental data were collected at $T$= 30 K and $T$ = 5 K in zero and high magnetic field of $H$ = 1.2 T, applied in-plane to the sample. We show the plots of specular reflectivities R$^{+}$ and R$^{-}$ ('+' and '-' correspond to neutron with spin parallel and anti-parallel to guide magnetic field) at $T$ = 5 K in $H$ = 0 T and $H$=1.2 T as well as at $T$ = 30 K in Fig. 3a-c. Unlike a ferromagnetic thin film or the nanostructured material where the reflectivities R$^{+}$ and R$^{-}$ are clearly separated from each other,\cite{Valeria,George} indicating the presence of net magnetic moment in the system, Nd thin film does not manifest such behavior (beyond two sigma of error bar) in $H$ = 0 T data at either temperature. It can occur due to two reasons: (a) either the system is non-magnetic and, hence, does not possess a net magnetic moment or, (b) the compensation of spin densities, as in the case of antiferromagnetic material, forbids the development of polarized moment, typically observable in PNR measurements. Nd is a known antiferromagnet,\cite{Bak} which can explain the absence of irreversibility between R$^{+}$ and R$^{-}$ curves in zero field. However, a small separation between R$^{+}$ and R$^{-}$ curves is observed in the PNR measurement in applied field, Fig. 3b, which indicates the development of field-induced magnetic moment in Nd film. The magnetization profile is uniform through the film thickness. So, no interface effect is detected.

The quantitative determination of the overall magnetization in applied field is obtained by analyzing the experimental results using a generic LICORNE-PY program,\cite{VL2,George} which generates the reflectivity pattern for a given set of physical parameters of the system e.g. layer thickness, density, interface roughness and magnetic moment of magnetic layer. The fitting was performed simultaneously to the data sets measured at different temperatures and fields. This procedure ensures the high accuracy of obtained parameters. Nuclear and magnetic scattering length densities (NSLD and MSLD), corresponding to the depth profiles of chemical structure and in-plane magnetization vector distributions, respectively, are depicted in Fig. 3d-f. The magnetization of Nd thin film is manifested by the spin asymmetry SA, given by (R$^{+}$ - R$^{-}$)/( R$^{+}$ + R$^{-}$). Experimental plot of spin asymmetry SA as a function of wave vector transfer Q at different temperatures and fields are shown in Fig. 3g-i. As expected, very small MSLD, corresponding to 1.2 emu/cc and 3.4 emu/cc are detected in experimental data at $T$ = 5 K and $T$ = 30 K in low field of 20 Oe, respectively. On the other hand, a MSLD with the net moment of 19.6 emu/cc is extracted from the field measurement data at $T$ = 5 K in Fig. 3h. Analysis of PNR data not only confirms the notion of field-induced magnetism in the Nd thin film, but also reveals the absolute value of magnetic moment at $H$ = 1.2 T.

\begin{figure}
\centering
\includegraphics[width=8.6cm]{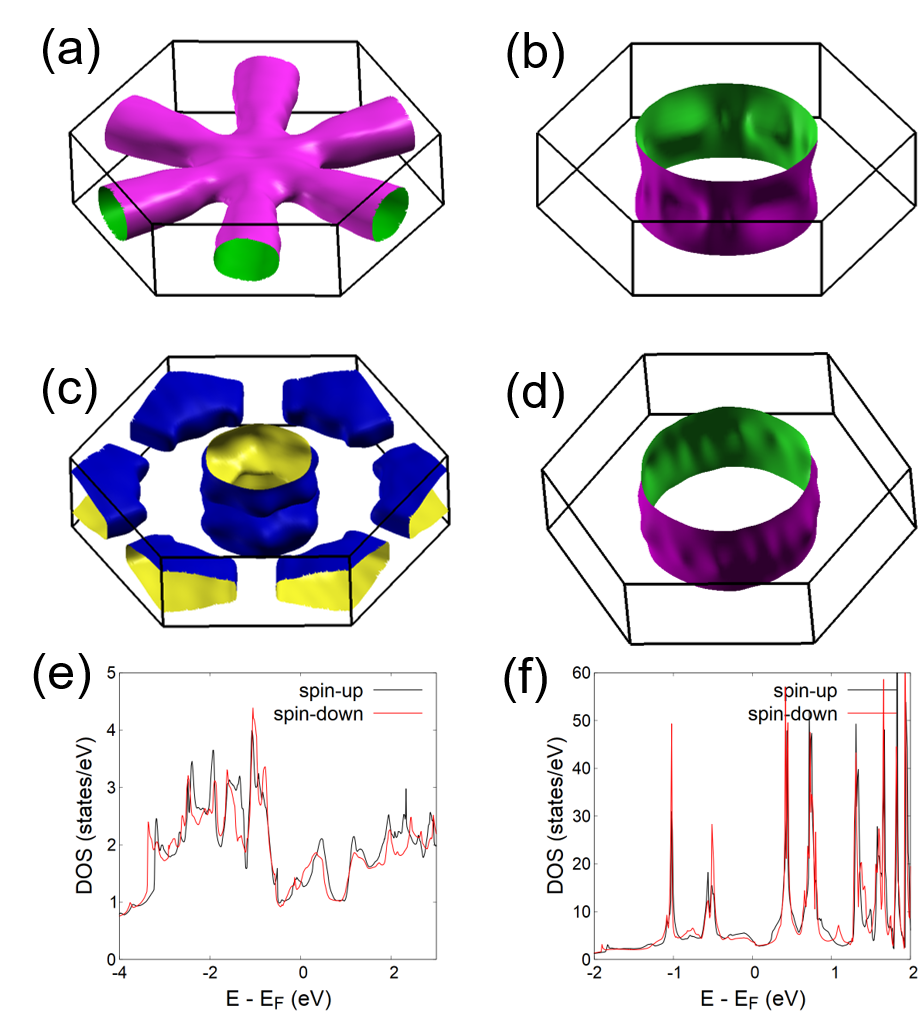} \vspace{-6mm}
\caption{(color online) Fermi surface properties of bulk and thin film specimens of Nd. (a-d) Spin resolved Fermi surfaces of bulk and thin-film of Nd, respectively: spin up (Fig. a, b) and spin down (Fig. c, d). (e-f) Density of states of bulk and thin-film of Nd, calculated using density functional theory (see text), elucidates the semiconducting character of the material. It is consistent with electrical measurements, exhibiting very weak metallic conductivity.} \vspace{-6mm}
\end{figure}

We have also tried to understand the electrical characteristics and the absence of magnetism in zero field in Nd film using the density functional theory (DFT) calculations. DFT calculations are performed for both the bulk and the thin film specimens of Nd. Although the study of bulk Nd is not the main focus of our study, it helps us in cross-checking the DFT results using the previously known parameters. DFT calculations are performed by employing the plane-wave basis set, as implemented in the Quantum-ESPRESSO.\cite{Baroni} The projector augmented wave method was used with Troullier-Martins norm-conserving pseudopotential with nonlinear core correction. The exchange correlation functional was treated within the generalized gradient approximation of Perdew-Burke-Ernzerhof (PBE-GGA).\cite{Perdew}  A well converged kinetic-energy cutoff of 60 Ry was used with a Monkhorst-Pack sampling of 16$\times$16$\times$4 (16$\times$16$\times$1) for the bulk (thin-film) case. Previously reported Nd lattice parameters were used as the initial configuration of the atoms for the bulk case.\cite{oldNd} For thin film calculations, we have used the supercell approximation and adopted the truncated Coulomb interaction method proposed in Ref.~\cite{Mauri}. For both bulk and thin film calculations, adjacent layers of Nd atoms were initialized to be antiferromagnetic. A strict self-consistent energy convergence criterion of 5$\times$10$^{-8}$ Ry is imposed. Fig.4 a, c show the calculated hole-type spin-resolved Fermi surfaces for the majority (spin-up) and minority (spin-down) carriers in the bulk case. One can see the difference in Fermi pockets which branches out of the highly-symmetric \textbf{k} points: M and K. In the case of the majority spin, as shown in Fig.4a, the Fermi surface extends from the M-point towards the $\Gamma$-point forming a centered pocket, whereas in the minority spin case, shown in Fig.4c, the Fermi pockets are clustered at the K-point and also centered at the $\Gamma$-point. On the other hand, the majority and minority Fermi surfaces in thin film, as shown in Fig.4b, d, manifest a truncated cylinder-shaped pocket centered at the $\Gamma$-point, which characterizes the presence of heavy carriers. As shown in Fig. 4e and 4f, the calculated spin-resolved density of states (DOS) in bulk Nd exhibits a weak splitting between the majority and minority states, whereas there is a close overlap in the case of the thin film. This indicates that the bulk Nd is antiferroagnetic in nature. The difference in electrical conductivity of the bulk Nd and the thin film Nd is manifested in the DOS at the Fermi level. In thin film Nd, the DOS at the Fermi level resembles the delta function and is much higher than that of the bulk Nd. It suggests the existence of a much larger carrier effective mass in the thin film case, which gives rise to much weaker electrical conductivity. This is in qualitative agreement with experimentally observed low electrical conductivity in the Nd thin film.

In summary, we have presented a comprehensive study of Nd thin film using electrical, magnetoresistance and polarized neutron reflectometry measurements. Our study elucidates important information regarding the electric and magnetic properties in Nd thin film that can play crucial role in the practical applications. Unlike bulk Nd, thin film of Nd is not very electrically conducive. At low temperature of 18 K $< T <$ 25 K, the resistivity registers a sharp increase due to the hopping of charge carriers between different sites. But it is not an insulator. This is a desirable property for many applications, such as in developing the high performance electrocatalyst. We have also shown that as grown Nd film does not possess net magnetization. In applied field, a weak magnetic moment develops at low temperature, $T <$ 18 K. The remnant characteristic of field induced magnetization causes the occurrence of competing quasi-stable states near zero field when magnetic field is reduced to $H$ = 0 T from a higher value. The experimental observation of the absence of magnetic moment in zero field is also confirmed by the DFT calculations. Future works on the study of Nd film grown on a different substrate, such as SrTiO$_{3}$, are highly desirable. Perhaps, a highly anisotropic substrate can cause an experimentally observable polarization of net magnetic moment in zero field.

DKS thankfully acknowledges the support by the Department of Energy, Office of Science, Office of Basic Energy Sciences under the grant no. DE-SC0014461. The
research conducted at the Spallation Neutron Source was sponsored by the Scientific User Facilities Division, Office of Basic Energy Sciences, U.S. Department of Energy.

\clearpage

\end{document}